\documentclass[12pt]{article}

\RequirePackage[numbers,compress]{natbib}
\usepackage[margin=2.0cm]{geometry}
\usepackage{graphicx}

\newcommand\pubdate{\today}

\def\Title#1{\begin{center} {\Large #1 } \end{center}}
\def\Author#1{\begin{center}{ \sc #1} \end{center}}
\def\Address#1{\begin{center}{ \it #1} \end{center}}

\newcommand\pubblock{\rightline{\begin{tabular}{l}  \\ 
         \pubdate  \end{tabular}}}
\newenvironment{Abstract}{\begin{quotation}  }{\end{quotation}}
\newenvironment{Presented}{\begin{quotation} \begin{center} 
             PRESENTED AT\end{center}\bigskip 
      \begin{center}\begin{large}}{\end{large}\end{center} \end{quotation}}

\usepackage{mathrsfs}
\usepackage{amssymb,bm}
\usepackage{amsmath}
\usepackage{mathtools}
\usepackage{physics}
\usepackage{slashed}
\newcommand{\nc}{N_\mathrm{c}}

\newcommand{\ft}{F_{2}}
\newcommand{\fl}{F_{\mathrm{L}}}
\newcommand{\nf}{n_\mathrm{f}}
\newcommand{\seq}{\sum_{q}^{\nf} e_q^2}

\newcommand{\cf}{C_\mathrm{F}}
\newcommand{\TR}{T_\mathrm{R}}
\newcommand{\seqav}{ \bar{e}_q^2}
\newcommand{\cgf }{ C}

\newcommand{\cflg }{ {C}_{\fl g}}
\newcommand{\cfls }{ {C}_{\fl \Sigma}}
\newcommand{\wft}{\widetilde{F}_2}
\newcommand{\cflwft }{ {C}_{\fl \wft}}
\newcommand{\wftp}{\widetilde{F'}_2}
\newcommand{\wfl}{\widetilde{F}_{\mathrm{L}}}
\newcommand{\wflp}{\widetilde{F'}_{\mathrm{L}}}
\newcommand{\wflpp}{\widetilde{F''}_{\mathrm{L}}}
\newcommand{\ftw }{F_2^{\rm W^-}}

\newcommand{\fk }{F_3}
\newcommand{\fkw }{F_3^{\rm W^-}}
\newcommand{\ftcw }{F_{2\rm c}^{\rm W^-}}

\newcommand{\xq }{ (x, Q^2)}
\newcommand{\eu }{ e_u^2}
\newcommand{\ed }{ e_d^2}
\newcommand{\es }{ e_s^2}
\newcommand{\as}{\alpha_\mathrm{s}}

\begin{document}
\begin{titlepage}
 \pubblock
\vfill
\Title{Collider physics with no PDFs}
\vfill

\Author{Tuomas Lappi, Heikki Mäntysaari, Hannu Paukkunen, and Mirja Tevio}
\Address{
Department of Physics, University of Jyvaskyla,  P.O. Box 35, 40014 University of Jyvaskyla, Finland
}
\Address{
Helsinki Institute of Physics, P.O. Box 64, 00014 University of Helsinki, Finland
}
\vfill
\begin{Abstract}
Measurements of Deep Inelastic Scattering (DIS) provide a powerful tool to probe the fundamental structure of protons and other nuclei. The DIS cross sections can be expressed in terms of structure functions which are conventionally expressed in terms of parton distribution functions (PDFs) that obey the DGLAP evolution equations. However, it is also possible to formulate the DGLAP evolution directly in terms of measurable DIS structure functions entirely sidestepping the need for introducing PDFs. We call this as the physical-basis approach. In a global analysis one would thereby directly parametrize the (observable) structure functions -- not the (unobservable) PDFs. Ideally, with data constraints at fixed $Q^2$, the initial condition for the evolution would be the same at each perturbative order (unlike for PDFs) and the approach thus provides a more clean test of the QCD dynamics. \\ \\
We first study a physical basis consisting of the structure functions $F_2$ and $F_{\rm L}$ in the fixed-flavour number scheme to the leading non-zero order in $\alpha_s$. We show how to express the quark singlet and gluon PDFs in terms of $F_2$ and $F_{\rm L}$ directly in momentum space which then leads to the DGLAP evolution of the structure functions $F_2$ and $F_{\rm L}$. In the second step we expand the physical basis to include six independent structure functions, which allows for a consistent global analysis. The steps towards NLO accuracy and the variable-flavour-number scheme are outlined. At NLO accuracy (when the scheme dependence of PDFs starts to play a part), we can take advatage of the physical basis and express e.g. the Drell-Yan cross sections at the LHC directly in terms of measurable DIS structure functions and thus without the scheme dependence.

\end{Abstract}
\vfill
\begin{Presented}
DIS2023: XXX International Workshop on Deep-Inelastic Scattering and
Related Subjects, \\
Michigan State University, USA, 27-31 March 2023 \\
     \includegraphics[width=9cm]{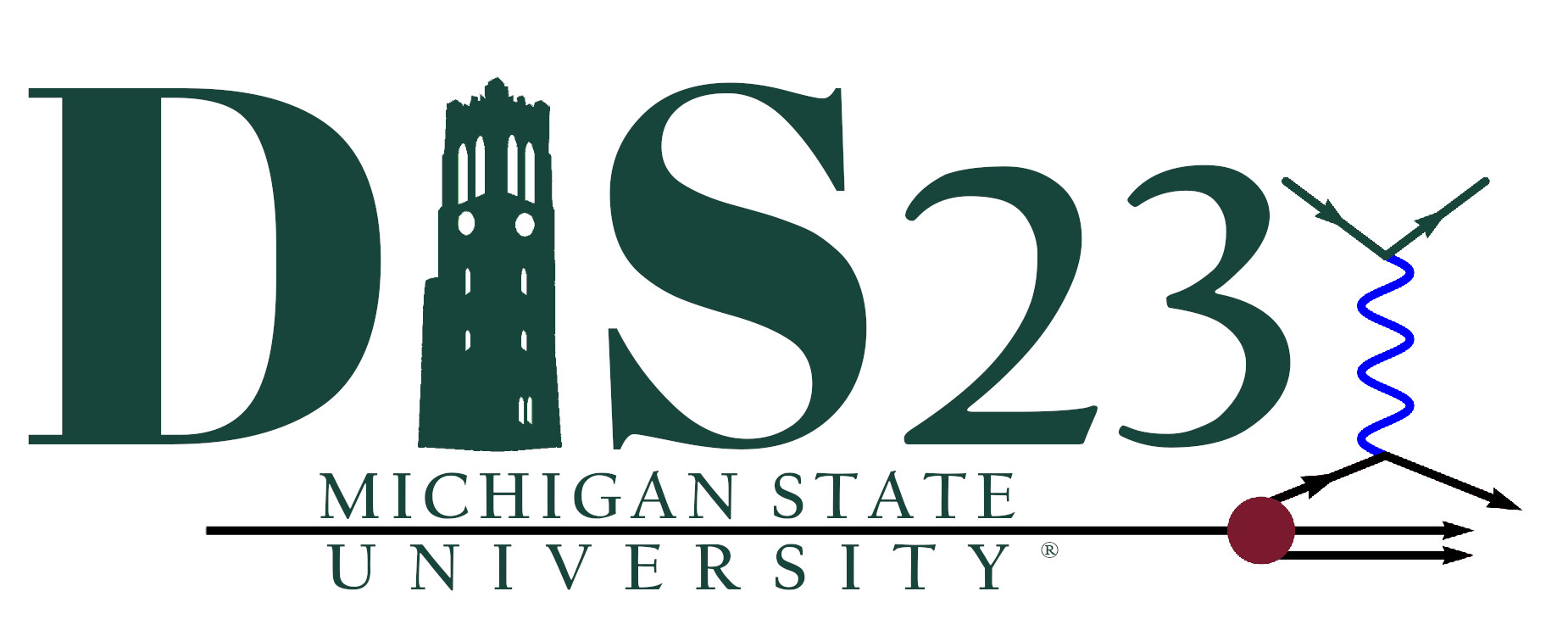}
\end{Presented}
\vfill
\end{titlepage}

\section{Introduction}
\addtocounter{page}{1}
As an alternative to conventional Dokshitzer-Gribov-Lipatov-Altarelli-Parisi (DGLAP) evolution for parton distribution functions (PDFs), one can also formulate the scale evolution directly for observable quantities such as the deep inelastic scattering (DIS) structure functions. The advantage of this approach is that then there is no need to define scheme-dependent renormalized parton distribution functions when going beyond the leading-order accuracy. 
In this work we will refer to this as the physical basis.
The existence of a physical basis follows from the fact that, as we explicitly demonstrate in this work, one can find a one-to-one mapping between the parton distribution functions and the DIS structure functions. Therefore  the parton distribution functions satisfying the DGLAP equation can also be expressed in terms of the DIS observables.
By introducing the DGLAP evolution in a physical basis one could -- in principle -- fix the initial condition for the evolution directly by data at fixed virtuality $Q^2$, and the intial condition would be the same to all orders in perturbation theory. In practice, however, the kinematic coverage of the available data is limited, and global fitting similar as in the case of PDFs will be necessary.

Conseptually physical basis is not a new idea, and it has been discussed for example in Ref.~\cite{Furmanski:1981cw,Catani:1996sc,Blumlein:2000wh,Hentschinski:2013zaa,Harland-Lang:2018bxd,Blumlein:2021lmf, vanNeerven:1999ca}. The novelty of our work~\cite{Lappi:2023lmi} is that, instead of studying a case specific physical basis, we construct a full dimension-six physical basis of structure functions which corresponds to PDFs with three active parton flavors. 
We use the first non-zero order in the running strong coupling $\as$ for $\ft$ and $\fl$ and solve the evolution equations directly in the momentum space. By formulating the final result in the momentum space we are able to analytically define the DGLAP evolution in physical basis in terms of observable structure functions only.

\section{Evolution in a two-observable physical basis}
\label{sec:1}
We first consider an illustrative approach to the physical basis where we have only two independent observables. In this case our physical basis consists of the DIS structure functions $\ft$ and $\fl$, taking into account only the massless quark singlet,
\begin{equation}
    \label{eq:singlet}
    \Sigma(x, \mu_f^2) = \sum_{q}\left[q(x, \mu_f^2)+\overline{q}(x, \mu_f^2)\right],
\end{equation} 
and the gluon PDF $g(x, \mu_f^2)$, where $\mu_f$ is the factorization scale. We work in the first non-zero order in $\as$, meaning that $\ft\sim \as^0$ and $\fl \sim \as^1$. The structure functions $\ft$ and $\fl$ can be written as
\begin{align}
    \frac{1}{\seqav} \frac{F_2(x, Q^2)}{x} & = C_{F_2\Sigma}^{(0)} \otimes \Sigma(x, \mu_f^2) \label{eq:F2master}, \\
    \frac{1}{\seqav} \frac{F_{\rm L}(x, Q^2)}{x} & = \frac{\as(\mu^2_r)}{2\pi} C_{F_{\rm L}\Sigma}^{(1)}  \otimes \Sigma(x, \mu_f^2) \label{eq:FLmaster}  + 2n_f \frac{\as(\mu^2_r)}{2\pi} C_{F_{\rm L}g}^{(1)} \otimes g(x, \mu_f^2)  \,,
\end{align}
where at the first non-zero order the coefficient functions are 
\begin{align}
\label{eq:C1F}
C_{F_2\Sigma}^{(0)}(z) =  \delta(1-z) \,, \quad
\cfls^{(1)}(z)  =  2 \cf z \,, \quad \text{and }
\cflg^{(1)}(z)  =  4 \TR z\left(1-z\right)\,. 
\end{align}
Here we used $N_c=C_A=3$, $\cf=(\nc^2-1)/(2\nc)$, and $\TR=1/2$. In this work $n_f=3$ is the number of massless flavours and $\seqav$ is the average quark charge,  
\begin{equation}
    \label{eq:eq2av}
   \seqav\equiv \frac{1}{\nf}\sum_q e_q^2 \,, 
\end{equation}
where $e_q$ denotes the electric charge of quark $q$.

The aim in this work is to write the $Q^2$ evolution equations for structure functions directly for the structure functions $\ft$ and $\fl$. This requires us to first invert Eqs. \eqref{eq:F2master} and \eqref{eq:FLmaster} such that the singlet and gluon PDF can be expressed in terms of $\ft$ and $\fl$. This results in
\begin{align}
    \Sigma(x,\mu_f^2) & = \frac{1}{\seqav} 
    \wft\xq
        \label{eq:sigmafromF2} \,, \\ 
    g(x, \mu_f^2) & = \frac{1}{\nf\seqav} \bigg( \cgf_{g\wftp}\otimes \wftp +\cgf_{g\wft}\otimes \wft +\cgf_{g\wflpp}\otimes\wflpp+\cgf_{g\wflp}\otimes\wflp+\cgf_{g\wfl}\otimes\wfl  \bigg) \,, \label{eq:gfromF2FL} 
\end{align}

where
\begin{align}
    \wft\xq  &\equiv \frac{\ft(x, Q^2)}{x} \qquad \qquad
    \wfl\xq \equiv  \frac{2\pi}{\as(\mu_r^2)} \frac{\fl(x, Q^2)}{x}  \\
    \widetilde{F'}_{2,L}\xq &\equiv  x\frac{\dd}{\dd{x}}\widetilde{F}_{2,L}(x, Q^2) \quad
    \wflpp\xq \equiv x^2\frac{\dd[2]}{\dd{x^2}}\wfl(x, Q^2) \,.
\end{align}
The coefficient functions $\cgf_{g\wftp}$, $\cgf_{g\wft}$, $\cgf_{g\wflpp}$, $\cgf_{g\wflp}$, and $\cgf_{g\wfl}$ are listed in Ref.~\cite{Lappi:2023lmi}.

When we take $Q^2$ derivatives of  Eqs.~(\ref{eq:F2master}) and (\ref{eq:FLmaster}), and then use Eqs.~(\ref{eq:sigmafromF2}) and (\ref{eq:gfromF2FL}), arrive with the evolution equations for $\ft$ and $\fl$,
\begin{align}
\label{eq:toyf2eq}
 \frac{\dd}{\dd\log Q^2} \Bigg[ \frac{F_2(x, Q^2)}{x} \bigg] & = 
      \frac{\as(Q^2)}{2\pi} \bigg[ C_{F_2\Sigma}^{(0)} \otimes P_{qq}  \otimes 
    \wft
             + 2  C_{F_2\Sigma}^{(0)} \otimes P_{qg} \otimes \bigg( \cgf_{g\wftp}\otimes \wftp\\
 & +\cgf_{g\wft}\otimes \wft +\cgf_{g\wflpp}\otimes\wflpp+\cgf_{g\wflp}\otimes\wflp+\cgf_{g\wfl}\otimes\wfl  \bigg) \Bigg] \nonumber \,, \\
 \frac{\dd}{\dd\log Q^2} \bigg[ \frac{2\pi}{\as(Q^2)} \frac{F_{\rm L}(x, Q^2)}{x} \bigg] & = 
 \left(\frac{\as(Q^2)}{2\pi}\right) \bigg[   C_{F_{\rm L}\Sigma}^{(1)} \otimes P_{qq} 
+  2n_f C_{F_{\rm L}g}^{(1)} \otimes P_{\rm gq} 
 \bigg] \otimes 
 \wft 
 \nonumber \\
& + 2 \left(\frac{\as(Q^2)}{2\pi}\right) \bigg[  C_{F_{\rm L}\Sigma}^{(1)} \otimes P_{qg} 
+ C_{F_{\rm L}g}^{(1)} \otimes P_{\rm gg}  \bigg] \\
 & \otimes \bigg( \cgf_{g\wftp}\otimes \wftp +\cgf_{g\wft}\otimes \wft +\cgf_{g\wflpp}\otimes\wflpp+\cgf_{g\wflp}\otimes\wflp+\cgf_{g\wfl}\otimes\wfl  \bigg) \nonumber \,,
\end{align}
where we have set the renormalization scale to be $\mu_r^2 = Q^2$. Here $P_{\rm qq}$, $P_{\rm qg}$, $P_{\rm gg}$, and $P_{\rm gq}$ are the LO splitting functions listed in Ref.~\cite{Lappi:2023lmi}.
These equations include double convolutions which can, however, be analytically reduced to a single one, see Ref.~\cite{Lappi:2023lmi}.

\section{Evolution of a six-observable physical basis}

In this section we repeat the same steps as in Sec. \ref{sec:1}, but now we consider a more complete setup by distinguishing between the light quark flavors. We still include only the light quark flavors and continue to work at the first non-zero order in $\as$. 

We separate the quark distributions $u \neq \overline{u}$, $d\neq \overline{d}$, but keep $s=\overline{s}$ to limit the number of observables  needed in the physical basis. Including the gluon distribution, we have in total six PDFs. In order to express the PDFs in terms of physical observables, we first need to collect a set of six linearly independent DIS structure functions. From neutral current DIS we choose the structure functions $\fl$, $\ft$ and $\fk$. From charged current DIS we choose $\ftw$, $\fkw$, and $\ftcw$ corresponding to the $W^{-}$-boson exchange.

In the first non-zero order in $\as$ structure functions $\ft$, $\fk$ $\ftw$, $\fkw$, and $\ftcw$ are expressed in terms of PDFs as~\cite{Furmanski:1981cw,ParticleDataGroup:2022pth,Moch:2004xu}
\begin{align}
\label{eq:F2-F3}
\left(\begin{array}{c} \ft \\
 \fk \\
 \ftw \\
 \fkw \\
 \ftcw \end{array} \right) & = 
 \left(
\begin{array}{ccccc}
x\ed & x\ed & x\eu & x\eu & 2x\es\\
2(L^2_d-R_d^2) & -2(L^2_d-R_d^2) & 2(L^2_u-R_u^2) & -2(L^2_u-R_u^2) &  0 \\
0 & 2x & 2x & 0 & 2x\\
0 & -2 & 2 & 0 & -2 \\
0 & 0 & 0 & 0 & 2x 
\end{array}
\right)
\left(
\begin{array}{c} d \\ \overline{d} \\ u \\ \overline{u} \\ \overline{s} \end{array} \right) .
\end{align}
 Here, $L_q = T_q^3-2e_q\sin^2 \theta_W$ and $R_q =-2e_q\sin^2 \theta_W$, where $\theta_W$ denotes the Weinberg angle and $T_q^3$ is the third component of the weak isospin. 
 Now the structure function $\fl$ reads
\begin{align}
     \label{eq:FLfullB}
    \fl(x, Q^2) = \frac{\as(Q^2)}{2\pi}x\left[\cflwft^{(1)}\otimes\wft(Q^2) +2\seq\cflg^{(1)}\otimes g(Q^2)\right] \,, 
\end{align}
where the coefficient functions $\cflwft^{(1)} =\cfls^{(1)}$ and $\cflg^{(1)}$ were defined in Eq. \eqref{eq:C1F}.
The expressions for PDFs in terms of the structure functions are listed in Ref.~\cite{Lappi:2023lmi}.

As in previous section, we can relate the DGLAP evolution of the structure functions to the DGLAP evolution of the quark and antiquark PDFs. By taking $Q^2$ derivatives of the structure functions defined in Eq.~\eqref{eq:F2-F3} we arrive at DGLAP evolutions for structure functions 
\begin{equation*}
    \ft,\quad \fk,\quad \ftw,\quad \fkw,\quad \ftcw,\quad \text{and } \fl,
\end{equation*}
which form a six dimensional physical basis.

\begin{figure}[t!]
  \includegraphics[width=.5\textwidth]{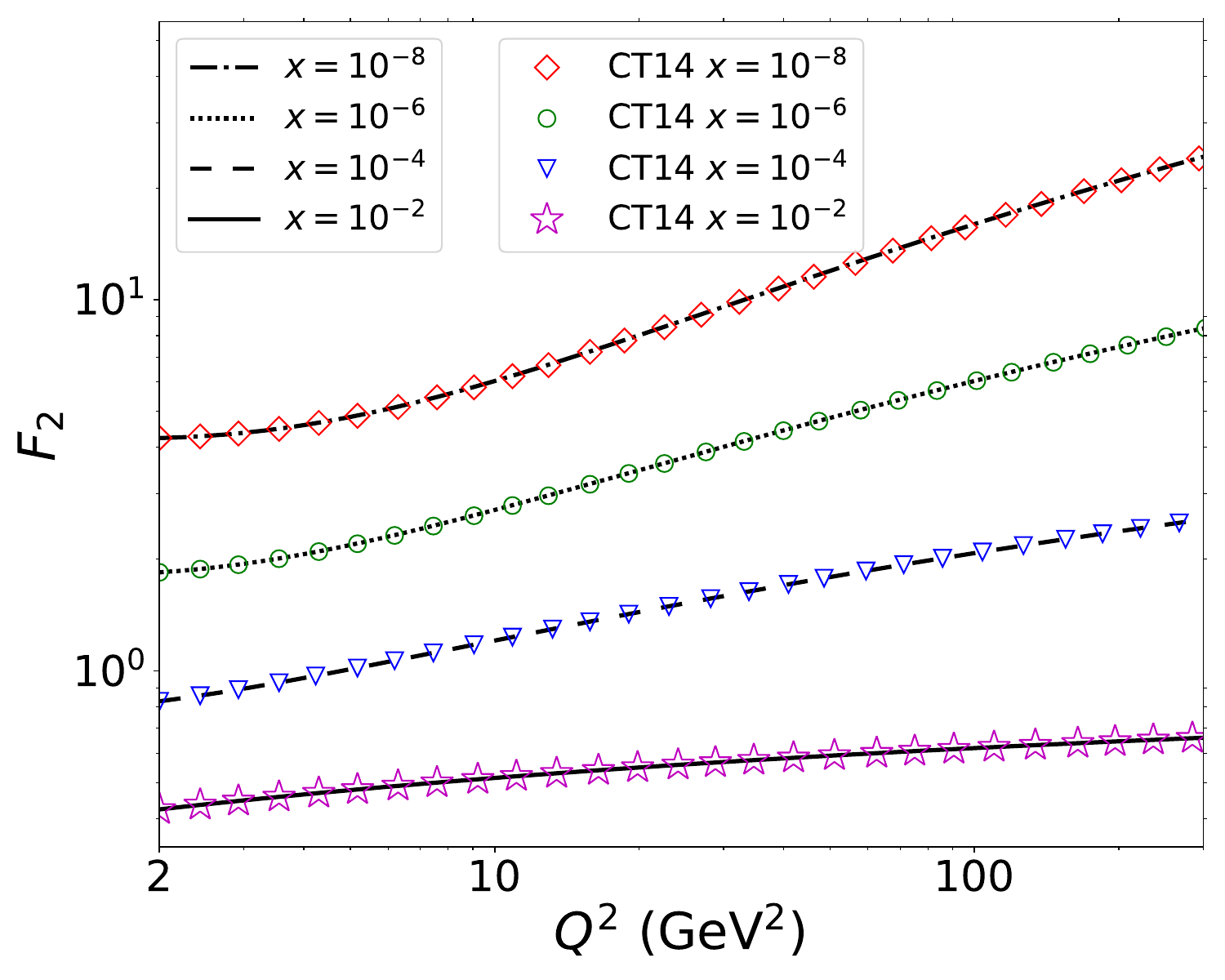}\hfill
  \includegraphics[width=.5\textwidth]{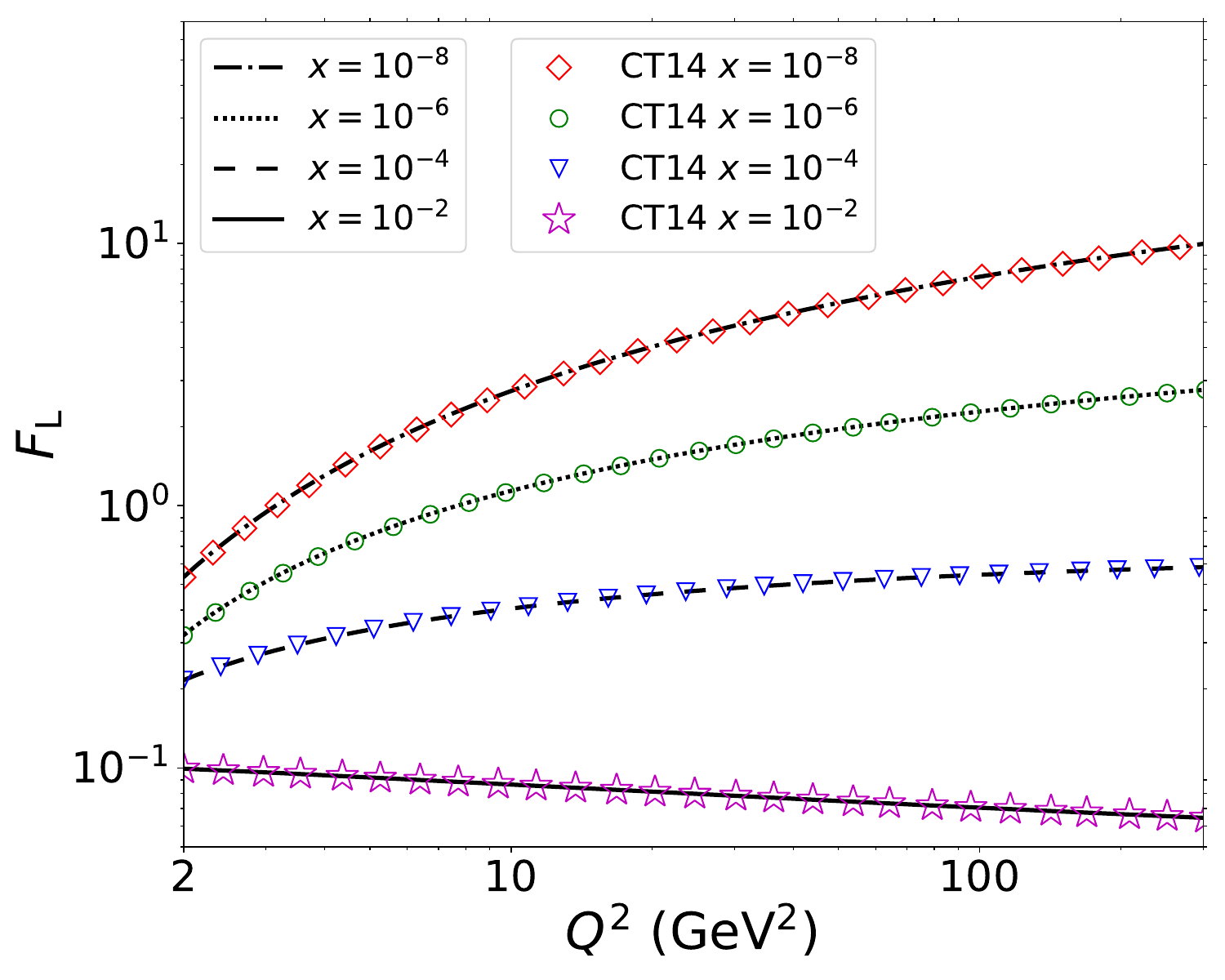}
  \includegraphics[width=.5\textwidth]{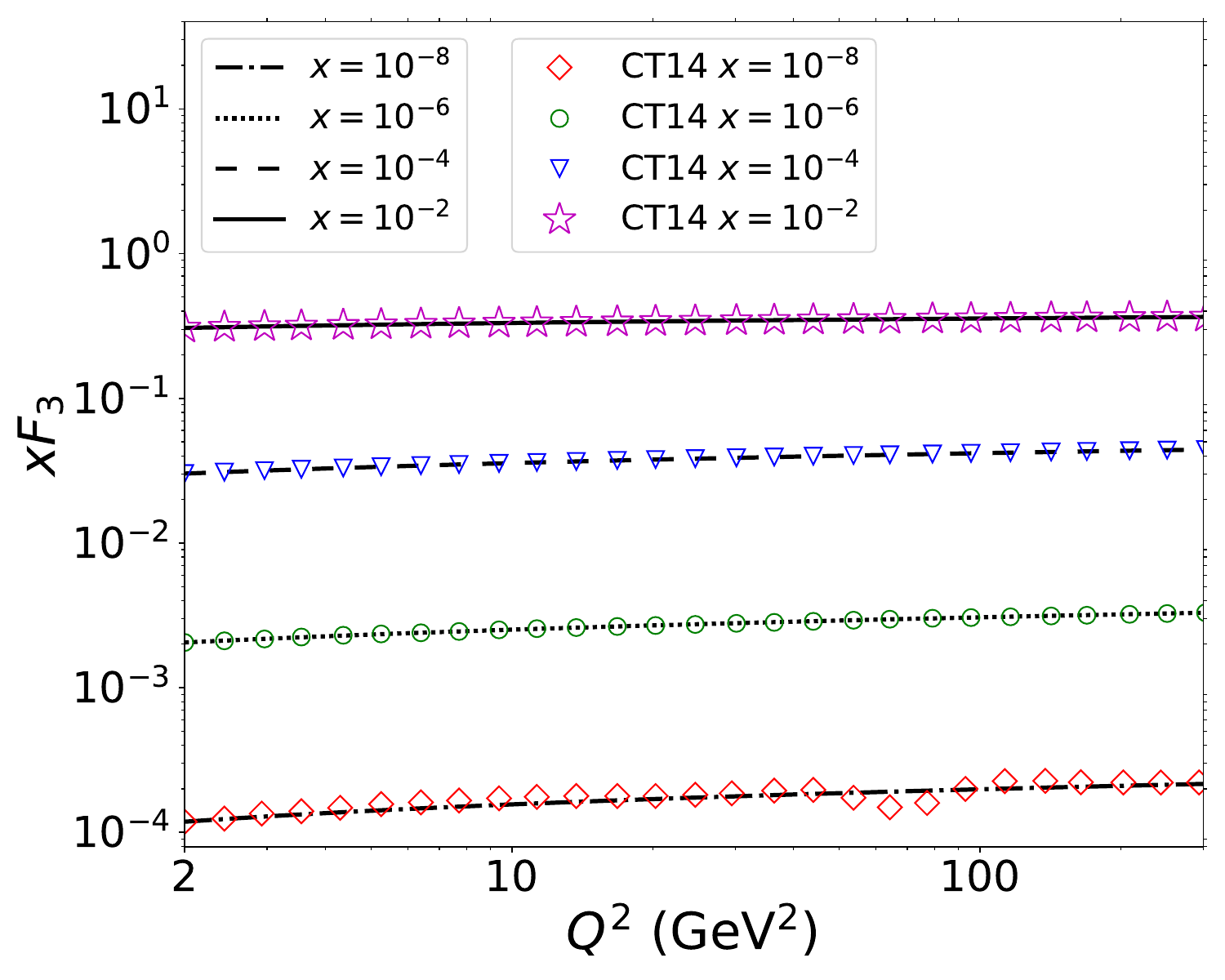}\hfill
  \includegraphics[width=.5\textwidth]{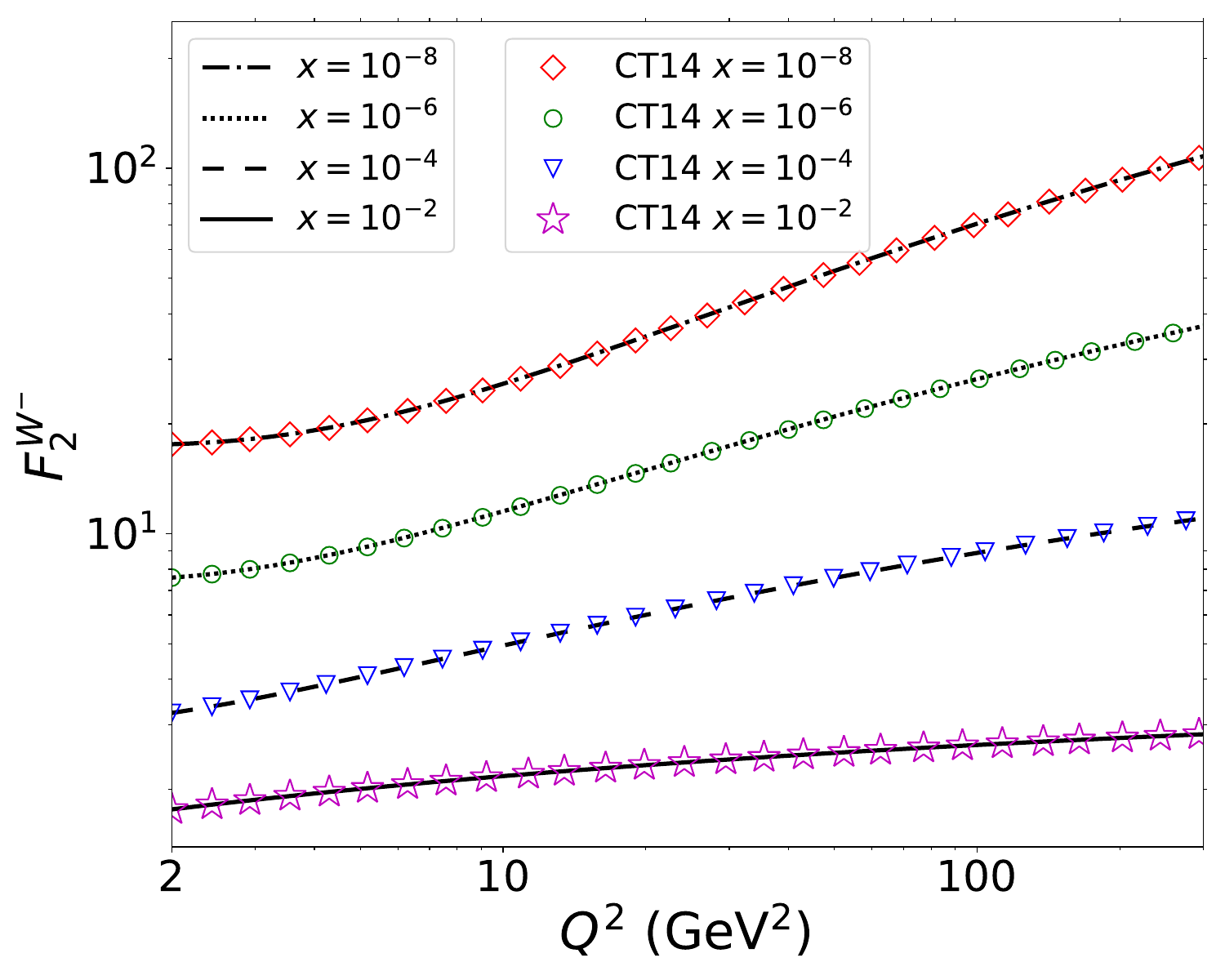}
  \caption{
  The $Q^2$ evolution of $\ft$, $\fl$, $\fk$, and $\ftw$ using the physical-basis approach (curves) compared with the usual PDF-based approach (markers).}
\label{fig:resultsFullBasis}
\end{figure}

The obtained $Q^2$ dependencies of $\ft$, $\fl$, $\fk$, and $\ftw$ in the six-observable physical basis are shown in Fig.~\ref{fig:resultsFullBasis}. Initial conditions for the physical basis evolution are computed at $Q^2 = 2.0 \, {\rm GeV}^2$  using Eqs.~\eqref{eq:F2-F3} and \eqref{eq:FLfullB} with the CTEQ (\texttt{CT14lo\_NF3}~\cite{Dulat:2015mca}) set of LO PDFs.
As expected, within the numerical accuracy, the $Q^2$ dependencies are found to match with the values obtained by computing the structure functions directly from Eqs.~\eqref{eq:F2-F3} and \eqref{eq:FLfullB} using DGLAP-evolved PDFs. The discrepancies for $\fk$ around $x=10^{-8}$ are presumably due to numerical noise. The overall excellent agreement validates the obtained evolution in the physical basis.

\section{Summary}

We have shown how the DGLAP evolution can be directly formulated in terms of observable DIS structure functions in the case of three light quarks, first non-zero order in $\as$. 
We first considered a toy model with only the light quark singlet and the gluon PDF, and constructed a physical basis with only structure functions $\ft$ and $\fl$. Then we proceeded to study a more complete case by considering the quark PDFs $u$, $\overline{u}$, $d$, $\overline{d}$, and $s=\overline{s}$ together with the gluon PDF. We constructed a corresponding physical basis with six observables $\fl$, $\ft$, $\fk$, $\ftw$, $\fkw$, and $\ftcw$.
We also confirmed numerically that the results obtained by performing the DGLAP evolution in the physical basis result in the same $Q^2$ evolution as in the conventional approach with DGLAP-evolved PDFs. 

In future work we will expand the perturbative order to reach the second non-zero order in $\as$. At that order the we can fully exploit the advantage of the physical basis advocated in this work, as it becomes possible to avoid the scheme dependence which otherwise manifests itself at NLO. 
In the future, we also intend to extend the procedure discussed in this work to cover the heavy quark flavors, and thus obtain a physical basis with more degrees of freedom.

\section*{Acknowledgements}
This work was supported under the European Union’s Horizon 2020 research and innovation programme by the European Research Council (ERC, grant agreement No. ERC-2018-ADG-835105 YoctoLHC) and by the STRONG-2020 project (grant agreement No. 824093). 
This work was also supported by the Academy of Finland, the Centre of Excellence in Quark Matter (projects 346324 and 346326), projects 321840 (T.L, M.T), project 308301 (H.P., M.T), and projects 338263 and 346567 (H.M).Views and opinions expressed are however those of the authors only and do not necessarily reflect those of the European Union or the European Research Council Executive Agency. Neither the European Union nor the granting authority can be held responsible for them.

\bibliographystyle{JHEP-2modlong}
\bibliography{refs}

\end{document}